\documentclass{revtex4}
\usepackage{epsfig}
\usepackage{graphicx}
\usepackage{amssymb}
\usepackage{amsmath}
\usepackage{amsthm}
\usepackage{setspace}
\usepackage[hypertex]{hyperref}
\usepackage{subfigure}
\def\complex{\mathbb{C}}
\newcommand{\CC}{\mathbb{C}}
\newcommand{\RR}{\mathbb{R}}

\newcommand{\coleq}{\mathrel{\mathop:}=}

\newtheorem{theorem}{Theorem}

\newtheorem{lemma}[theorem]{Lemma}

\theoremstyle{definition}
\newtheorem{definition}[theorem]{Definition}

\newcommand*{\cA}{\mathcal{A}} %
\newcommand*{\cB}{\mathcal{B}}

\newcommand*{\cH}{\mathcal{H}}
\newcommand*{\cI}{\mathcal{I}}

\newcommand*{\cL}{\mathcal{L}}

\newcommand*{\cX}{\mathcal{X}}

\newcommand*{\tr}{\mathsf{Tr}}

\newcommand{\be}{\begin{align}}
\newcommand{\ee}{\end{align}}
\newcommand{\bea}{\begin{eqnarray}}
\newcommand{\eea}{\end{eqnarray}}
\newcommand{\bestar}{\begin{equation*}}
\newcommand{\eestar}{\end{equation*}}
\newcommand{\beastar}{\begin{eqnarray*}}
\newcommand{\eeastar}{\end{eqnarray*}}

\def\one{\leavevmode\hbox{\small1\normalsize\kern-.33em1}}

\newcommand{\boxy}{\square}
\newcommand{\tmin}{\otimes_\text{min}}
\newcommand{\tmax}{\otimes_\text{max}}

\begin{document}
\title{Finite de Finetti theorem for conditional probability distributions describing physical theories}

\author{Matthias Christandl}

\email{christandl@lmu.de}
\affiliation{Centre for Quantum Computation, Department of Applied
  Mathematics and Theoretical Physics, University of Cambridge,
  Wilberforce Road, Cambridge CB3~0WA, United Kingdom}
\affiliation{Arnold Sommerfeld Center for Theoretical Physics, Faculty of Physics, Ludwig-Maximilians-Universit{\"a}t M{\"u}nchen, Theresienstrasse 37, 80333 Munich, Germany}

\author{Ben Toner}
\email{bentoner@bentoner.com}
\affiliation{School of Physics, The University of Melbourne, Victoria 3010, Australia}
\affiliation{Centrum voor Wiskunde en Informatica, Kruislaan 413, 1098
  SJ Amsterdam, The Netherlands}
\affiliation{Institute for Quantum Information, California Institute of Technology, Pasadena CA 91125, USA}

\begin{abstract}
We work in a general framework where the state of a physical system
is defined by its behaviour under measurement and the global state
is constrained by no-signalling conditions. We show that the marginals of symmetric states in such theories can
be approximated by convex combinations of independent and identical
conditional probability distributions, generalizing the
classical finite de~Finetti theorem of Diaconis and Freedman. Our results apply to
correlations obtained from quantum states even when there is no
bound on the local dimension, so that known quantum de~Finetti
theorems cannot be used.
\end{abstract}
\maketitle

\section{Introduction}
Given a bowl containing $n$ colored balls, we wish to compare two ways of
obtaining a random sample of $k\leq n$ balls: (i) we randomly choose a ball,
replace it with a ball of the same color, and repeat this step $k$ times; (ii)
we do the same but don't replace the balls. If $k \ll n$, then the probability
of obtaining a particular set of $k$ balls will be almost the same in both
cases~\cite{diaconis80:_finit}. This observation has profound consequences for
Bayesian statistical inference, as we now describe.

Suppose we perform an experiment $k$ times in order to estimate some physical
quantity, e.g., the probability $\lambda$ that a muon decays in
a given time. Let $A_i =1$ if the $i$th muon decayed and $A_i=0$ if it did not. If we
assume that the results of the experiments are independent, we can posit some prior probability
distribution $m(\lambda)$  and analyze our data by updating this probability
distribution as more data arrives. Statisticians of de Finetti's
subjective school~\cite{bernardo94:_bayes_theor} are not willing to accept this assumption, however, since for
them all probability distributions should be subjective degrees of belief, which $m(\lambda)$ is not.
Instead, they make the weaker assumptions that the experiment could have
been performed $n \gg k$ times and that there was nothing special about the
experiments actually performed. These assumptions, together with the observation about colored balls above,
can be shown to imply that there exists a distribution $m(\lambda)$ such that
\begin{align}  P[A_1, \ldots, A_k] \approx \int dm(\lambda)
  P_\lambda[A_1]\cdots P_\lambda[A_k], \end{align} i.e., the probability
distribution $P[A^k]$ behaves \emph{as if} the experiments really were
independent and there really were some objective prior $m(\lambda)$.  This is a
statement of the famous \emph{de Finetti representation
  theorem}~\cite{finetti37:_la,diaconis80:_finit}. Our results establish the
same correspondence for measurement results in a more general, probabilistic, physical theory, where the state of a system is described by a conditional probability distribution.
%(see~\cite{Barnum:06,Barnum:07}).

%Our work Recently, quantum-mechanical analogues of the de Finetti theorem have been studied and found far-reaching applications in quantum cryptography and quantum information theory in general~\cite.

We now give a brief description of the setting and our results; precise definitions are given later on. A physical system in a probabilistic physical theory is made up of a number of---in our case
identical---subsystems, called \emph{particles}. On each particle
different measurements from a set $\cX$ can be performed and outputs
from a set $\cA$ are obtained. The state of
a particle is specified by a \emph{conditional probability
  distribution} $P[A|X]$: the probability of obtaining result $a$ when
performing measurement $x$ is given by $P[A=a|X=x]$. The possible states of $n$ particles are the conditional probability distributions $P[A^n|X^n]$ that obey a \emph{no-signalling} property, which ensures that the reduced
state on a subset of the particles is always well-defined.%

Our main result is that the joint state $P[A^k|X^k]=P[A_1 \cdots A_k|X_1 \cdots X_k]$ of $k$ particles randomly chosen
from $n$ particles---or equivalently, the state of the first $k$ particles of a permutation-invariant state of $n$ particles---can be approximated by a convex
combination of identical and independent conditional probability
distributions,
\begin{align} \label{box-approx} P[A^k|X^k] \approx \int dm(\lambda) P_\lambda[A|X]^{\times k}  \end{align}
and that the error in the
approximation is bounded by $|\cX|k(k-1)/n$ in the appropriate distance measure, where $|\cX|$ is the number of different possible measurements. %~\footnote{Individual particles are labeled by subscripts and $m(\lambda)$ is a measure on the set of conditional probability distributions on a single particle.}.
(We write $P_\lambda[A|X]^{\times k}$ for $P_\lambda[A_1|X_1] \cdots P_\lambda[A_k|X_k]$.)
 Our result generalizes
the finite de Finetti theorem of Diaconis and Freedman, who proved for classical probability distributions ($|\cX|=1$) that the error in the approximation is no more
than $k(k-1)/n$~\cite{diaconis80:_finit}~\footnote{
Diaconis and
Freedman also obtain a second bound $ k|\cA|/n$. The analogous bound
within our framework is $k |\cA|^{|\cX|}/n$. Restricting
to \emph{adaptive} measurements on individual particles, we are able to improve this bound
to $k |\cX|^2|\cA|(1+4\sqrt{\frac{2+\log |\cX|}{k}})/n $~\cite{benmat:long}.}.

This paper is motivated by recent work on finite \emph{quantum} de Finetti
theorems, i.e., statements of the form
\begin{align}
\label{quantum-approx}\rho^k \approx \int d\sigma \ \sigma^{\otimes
k},
\end{align} where $\rho^k$ is the $k$-particle reduced density matrix of a
permutation-invariant density matrix of $n$ particles with state space of dimension $d$, where the
error is at most $4d^2k/n$ in the trace distance~\cite{Koenig:04a,
  christandl06:_one}~\footnote{ A density operator $\rho^n$ is
  permutation-invariant if $\rho^n=\pi \rho^n \pi^{-1}$ for all permutations
  $\pi \in S_n$.}. In fact, it is necessary that the error depends on $d$~\cite{christandl06:_one}, and so the quantum de Finetti is not useful in applications where $d$ cannot be bounded.
Our results are designed to apply in this setting: provided we have a bound on the number
of ways $|\cX|$ that a system is measured, the approximation in Eq.~\eqref{box-approx} will be good, even if there is no bound on the local dimension $d$.
% One example of when we do not want to assume a bound on the local dimension
% is in cryptography, if the systems are prepared by an adversary. The quantum
% de Finetti theorem, and especially Renner's so-called `exponential' de
% Finetti theorem~\cite{renner05:_secur,renner:nature}, are integral in the
% analysis of quantum key distribution
% schemes~\cite{Bennett:84a,PhysRevLett.67.661}.
% (see~\cite{Barrett:05a,acin05:_from_theor_secur_quant_key_distr,scarani:_secrec,masanes:_uncon})
In recent years, quantum de Finetti theorems, especially Renner's so-called
`exponential' version~\cite{renner05:_secur}, have been used to
prove the security of quantum key distribution (QKD)
schemes~\cite{QKD}. At the same time, attempts have been
made to lift the assumption of a fixed (finite) local
dimension~\cite{acin:230501}.  Since quantum de
Finetti theorems are necessarily dimension-dependent, they cannot be used in
this setting. Although our theorems do not directly lead to security proofs either, we regard them as a first step towards this goal.

% Our results also apply to quantum systems, in the setting
% where there is a symmetric quantum state on $n$ systems and we make local
% measurements on $k$ of them.

We also prove a finite quantum de Finetti theorem for separable $\rho^n$: in
this case there is an approximation of the form in Eq.~\eqref{quantum-approx}
with error $k(k-1)/n$, \emph{independent} of the dimension. We do not, however, know
whether our techniques can be extended to prove the finite quantum de Finetti
theorem in full generality.  The issue is that our theorem concerns
conditional probability distributions that arise from measuring quantum
states and not the quantum states themselves. If we take, for example, a
tomographically complete set of measurements, the representation described in
Eq.~\eqref{box-approx} will in general contain distributions $P_\lambda[A|X]$ that cannot be obtained by
performing the tomographic measurements on quantum states. One can, however,
apply the argument of~\cite{caves:4537} to obtain the
infinite quantum de Finetti theorem and indeed an infinite de Finetti theorem for
any physical theory in what is known as the \emph{convex sets framework}~\cite{Barnum:06,Barnum:07} (see~\cite{BarrettLeifer06} for the details).

Another application of our work is to the study of classical channels.
% Up to now we have
% described de Finetti theorems as bounding the distance between states in a physical theory, and have not addressed
% operations or channels on those states.
Fuchs, Schack and
Scudo have used the Jamiolkowski isomorphism to transfer the infinite
quantum de Finetti theorem ($n=\infty$, $k<\infty$)~\cite{stormer69:_symmet_states_infin_tensor_produc_c,caves:4537} to
quantum channels~\cite{fuchs:062305}. Since a conditional probability distribution can be viewed as
a classical channel with probability distributions as input and
output, our results also provide a de Finetti theorem for
classical channels.

\emph{Outline.}---Our first task is to define an appropriate distance measure on
states of $k$ particles in probabilistic theories, in order to quantify the error in
Eq.~\eqref{box-approx}. The distance between states should bound the
probability of distinguishing them by measurement, and so we need to
be clear about what measurement strategies are allowed. One
possibility, which we explore in~\cite{benmat:long}, is to restrict
to strategies where each of the $k$ particles is measured
individually. But when the conditional probability distributions arise
from making informationally complete local measurements on entangled quantum states, the
resulting distance measure fails to bound the trace distance between
the quantum states. In the next section we show how to define a `good' distance measure in which all noncontextual
measurements are allowed, including all joint quantum-mechanical
measurements. We then state and prove our results. In the last section, we explain the origin of the distance measure, the convex sets framework, which allows us to conclude with an open question on finite de Finetti theorems in this more general setting.

\section{A distance measure for conditional probability distributions}
\label{sec:dist-meas-cond}

When we measure a quantum system, the probability of obtaining an
outcome $a \in \cA$ depends on which measurement $x \in \cX$ we choose
to perform on the system.  It is usual to describe a quantum system
using the formalism of density matrices, Hilbert spaces, and so on,
but we can also describe the system by specifying a conditional
probability distribution $P[A|X]$, where we write $P[A|X=x]$ for the distribution of measurement outcome $A$ given that measurement $x$ is performed~\footnote{In quantum theory, the most
general measurement is termed a positive operator-valued measure
(POVM).
If we perform a POVM $x$ with effects $E_{x,a}$
(satisfying $E_{x,a}= E_{x,a}^\dagger$, $E_{x,a}\succeq 0$ and $\sum_a E_{x,a} = \one$) on a system in state $\rho$, then the
distribution of the measurement outcome $A$ is given by
$P[A=a|X=x] = \tr \left(\rho E_{x,a}\right)$.}.
While a classical system can be described using an
\emph{unconditional} probability distribution, the same is not true
for a quantum system, since measuring a quantum system disturbs it,
eliminating our ability to make a second, incompatible, measurement on
the same system.
%Measuring a quantum system disturbs it, eliminating our ability to
%make a second, incompatible, measurement on the same system.  We can
%describe the results of measuring a quantum system via a conditional
%probability distribution for the measurement outcomes, conditioned
%on which measurement we choose to perform.  The need for the
%probability distribution to be conditional does not arise
%classically, where all measurement are compatible.

We are therefore motivated to describe the state of an abstract system
(not necessarily obeying quantum theory) using a conditional
probability distribution $P[A|X]$.  We view the
conditional probability distribution $P[A|X]$ as the output
distribution of a measurement that has been performed on system $A$.
Alternatively, one can view $P[A|X]$ as a channel that produces an
output distribution $P[A|X=x]$ on input $x$.  For this reason we refer
to the measurement setting $x$ as the \emph{input} and the measurement
result $a$ as the \emph{output}.
%Again we view $X$ as the
%label of the system and not as a random variable.
%Of course, one can
%introduce a probability distribution on $X$, say $Q[X]$, so that
%$P[A|X]Q[X]$ is a probability distribution on the combined system
%$AX$.
Generalizing from conditional probability distributions of one system, we
shall consider a conditional probability distribution $P[A^n|X^n]=P[A_1\cdots A_n|X_1 \cdots X_n]$, which
describes an abstract system composed of $n$ subsystems, which we call particles.
%For convenience, we shall speak of the subsystems as if they are held by $n$ separate \emph{parties} at $n$ separate \emph{sites}.

We need to be able to describe the state of a subset $\cI \subset \{1,
\ldots, n\}$ of the particles. Taking the marginal of a conditional
probability distribution $P[A^n|X^n]$ yields a conditional
distribution $P[A_{\cI}|X^n]$, where the outputs at the particles in
$\cI$ depends on the inputs at all $n$ sites.  In order to trace out
the particles that are not in $\cI$ entirely, rather than just the
outputs obtained from measuring them, we need another notion, that of
a conditional probability distribution being \emph{no-signalling}.
\begin{definition}
A conditional distribution $P[A^n|X^n]$
is \emph{no-signalling} if for all subsets $\cI \subset \{1, \ldots, n\}$ with complements $\bar\cI:=\{1, \ldots, n\} \backslash \cI$
\begin{equation}
\label{eq:10}
P[A_{\cI} = a_{\cI} |X_{\cI} = x_{\cI}] \coleq \sum_{a_{\bar{\cI}}} P[A^n = a^n|X^n=x^n]
\end{equation}
is independent of $x_{\bar\cI}$ for all $a_{\cI}$ and all $x_{\cI}$.
\end{definition}
The terminology derives from the following fact:
if we divide the $n$ parties into two groups, $\cI$ and $\bar\cI$, then, provided $P[A^n|X^n]$ is
no-signalling, it is impossible for the group $\cI$ to send
a signal to the group of $\bar\cI$ just by changing their inputs. Not
all conditional probability distributions are
no-signalling; for example, $P[A_1=a_1,A_2=a_2|X_1=x_1,X_2=x_2] = [a_1=x_2][a_2=x_1]$ (where $[t]$ is $1$ if $t$ is true and $0$ otherwise) is signalling.  We note that any conditional
probability distribution that arises from making local measurements on
a quantum state is no-signalling. The no-signalling requirement is the minimal assumption necessary to ensure that state of any subset of particles is well-defined. 

The goal of this paper is to approximate by product distributions a no-signalling conditional probability distribution on $k$ particles arising from a symmetric conditional probability distribution on $n$ systems, so we need to introduce a notion of distance for conditional probability distributions.  This distance measure should generalize the classical variational
distance, which is equal to the maximum probability of distinguishing two probability
distributions, and the quantum trace distance, which is equal to the maximal probability
of distinguishing two quantum states. In order to define a \emph{trace distance} for no-signalling conditional probability distributions we therefore need to determine
what measurement strategies can be used to distinguish two conditional
probability distributions.  In fact, there are three natural sets of measurement
strategies for conditional probability distributions, each of which
induces a distance measure on conditional probability distributions. We will work with the largest of these sets giving the strongest notion of a distance, for if we can show that two conditional probability distributions are almost indistinguishable using a particular set of measurements, it will trivially follow that they are also almost indistinguishable when only a subset of those measurements is allowed. Let us start by introducing the three
sets.

An \emph{individual measurement} is a distribution $P[X^k]$ on the inputs that maps the conditional probability distribution to the {unconditional} probability distribution $P[A^kX^k] =
P[A^k|X^k]P[X^k]$.  Such a measurement can be carried out by measuring each subsystem
individually.  Note that individual measurements also make sense if we drop the condition that $P[A^n|X^n]$ is no-signalling. Since we restrict to no-signalling conditional probability distributions, a
larger class of measurements is possible and indeed needed for
applications.  Suppose the conditional distribution $P[A_1A_2|X_1X_2]$ is no-signalling. We start by writing
\begin{align}
P[A_1A_2|X_1X_2=x_1x_2]&=P[A_1|X_1X_2=x_1x_2]P[A_2|A_1,X_1X_2=x_1x_2]\\
&=P[A_1|X_1=x_1]P[A_2|A_1,X_1X_2=x_1x_2],
\end{align}
where we made use of the no-signalling principle, Eq.~(\ref{eq:10}),
in the second line. This provides an operational means to sample from
$P[A_1A_2|X_1X_2=x_1x_2]$: We first sample $a_1$ from the distribution
$P[A_1|X_1=x_1]$, then sample $a_2$ from
$P[A_2|A_1=a_1,X_1X_2=x_1x_2]$.  The important point is that a no-signalling conditional probability distribution can provide
the output on system 1 before specifying which input is chosen for
system 2.  Therefore the
following \emph{adaptive measurement} on $P[A_1A_2|X_1X_2]$
is possible: Input $x_1$, obtain $a_1$, and choose an input $x_2=f(a_1)$,
where $f: \cA \to \cX$ is an arbitrary function. Such a strategy can lead to a higher probability of distinguishing two no-signalling conditional probability distributions, compared to individual
strategies~\footnote{Let $\cA = \cX = \{0,1\}$ and define
$P[A_1A_2=a_1a_2|X_1X_2=x_1x_2] = \frac12\, [a_1 + a_2  = x_1 \and x_2 \pmod 2].$
This
distribution is known as a nonlocal box~\cite{Popescu:94a} and one can easily check that it is
no-signalling.  We wish to distinguish this distribution from the  distribution
$Q[A_1A_2|X_1X_2]$, defined by
$ Q[A_1A_2=a_1a_2|X_1X_2 = x_1x_2] = \frac{1}{2}\, [a_2=1].$
(This is an unconditional product distribution where the first bit is random and the second bit is always one.)
For every setting of $x_1$ and $x_2$, $P[A_2 = 1|X_1X_2=x_1x_2] = 1/2$, and thus $P$ and $Q$ cannot be perfectly distinguished by making a measurement on both systems in parallel. But if we allow adaptive
strategies, then we can distinguish $P$ and $Q$ perfectly.  For instance, set $x_1 = 1$ and then set $x_2=a_1$, so that we have $a_1+a_2 = 1.a_1 \pmod 2$ and it
follows that $P[A_2=0] =1$.  Since $Q[A_2=0]=0$, we conclude that we can distinguish $P$ and $Q$ perfectly.}.

As in most of the paper we draw intuition from
quantum-mechanical correlations.  It is a well-established
fact that the distinguishability of quantum states depends on
whether individual or adaptive measurement strategies are
considered.  In the quantum case, furthermore, it is possible to
apply a joint measurement to all $k$ systems at once, a class of measurement which strictly contains adaptive
 measurements and can lead to strictly higher
distinguishability. \emph{Quantum data hiding} is an important application of this phenomenon~\cite{EggWer2002,hayden:062339}.

In defining joint operations on no-signalling conditional probability distributions, we essentially wish to allow all possible measurements whose outcomes behave like probability distributions. Motivated by this, we think of a no-signalling conditional probability distribution $P[A^k|X^k]$ as a vector in a real $|\cA|^k|\cX|^k$-dimensional space and consider linear functions from this space to a real ${|\cA|^k}$-dimensional space. The set of \emph{general measurements} is the set of linear functions $M$ such that $M(P[A^k|X^k])$ is a probability distribution for all no-signalling conditional probability distributions $P[A^k|X^k]$. Clearly, individual and adaptive strategies belong to the set of general measurements, but it includes strictly more strategies, too. (The assumption of linearity is necessary so that our probability behave reasonably when we take convex combinations of states and measurements; see Ref.~\cite{Barrett:05}.) 

\begin{definition}
\label{sec:introduction-general}\label{def:tracedistance}
  The \emph{trace distance} between two no-signalling conditional probability distributions $P[A^k|X^k]$ and $Q[A^k|X^k]$ is given by
\begin{align}
\|P[A^k|X^k]- & Q[A^k|X^k]\| \coleq \sup_{M}
\|M(P[A^k|X^k])-M(Q[A^k|X^k]) \|.\label{eq:18}
\end{align}
where the supremum is taken over all general measurements and $\|R[B] - S[B]  \|$ is the classical variational distance for probability distributions $R[B]$ and $S[B]$ on system $B$. Extending the definition by imposing linearity, $|| \cdot ||$ is a norm on the space of (real) linear combinations of conditional probability distributions and hence obeys the triangle inequality.
\end{definition}

A theory in which conditional probability distributions describe the state of a particle and where joint states of particles obey a no-signalling distribution can be treated in the \emph{convex sets framework}. The distance measure we introduced arises naturally in this framework. We review the convex sets framework in Section~\ref{sec:conv-sets-fram}. This will give us a broader view on de Finetti theorems and will allow us to pose an open question regarding de Finetti theorems in the convex sets framework.

\section{Our results}
\label{sec:our-results}
Suppose we have a conditional probability distribution $P[A^n|X^n]$ describing $n$ particles. If we interchange the particles according to a permutation $\pi \in S_n$, the resulting conditional probability distribution is
\begin{align*} &\pi P[A^n=a_1 \cdots a_n| X^n=x_1 \cdots
x_n]\\ &=
P[A^n=a_{\pi^{-1}(1)} \cdots a_{\pi^{-1}(n)}| X^n=x_{\pi^{-1}(1)}
\cdots x_{\pi^{-1}(n)}].
\end{align*}
We say that a conditional probability distribution $P[A^n|X^n]$ is
\emph{symmetric} if it is invariant under all permutations $\pi \in S_n$.
  If
$|\cX| = 1$, this definition reduces to the usual definition of a
symmetric probability distribution.
We can now state our main result:

\begin{theorem}\label{theorem:1}
Suppose that $P[A^n|X^n]$ is a symmetric no-signalling conditional probability distribution. Then there exists 
a probability distribution $p_\lambda$ such that
\begin{align}
\begin{split}
\bigl\|P[A^k|X^k]- & \sum_\lambda p_\lambda P_\lambda[A|X]^{\times k} \bigr\|  \leq
 \min \biggl( \frac{2k|\cX||\cA|^{|\cX|}}{n}, \frac{|\cX|k(k-1)}{n}\biggr),
\end{split}
\end{align}
where the distribution $p_\lambda$ is on a finite set of single-particle conditional probability distributions, labeled by $\lambda$.
\end{theorem}

This establishes that the state of a random subset of $k$ out of $n$ particles is well approximated by a convex combination of independent and identically distributed conditional probability distributions.
To prove Theorem~\ref{theorem:1}, we first show that if $P[A^n|X^n]$ is symmetric and $m$
is chosen to be sufficiently small, then $P[A^m|X^m]$ is separable (Lemma~\ref{lemma:product}). We then establish a de Finetti theorem for separable states, Lemma~\ref{th:convex}, which will complete the proof of our main result, Theorem~\ref{theorem:1}. We continue with Lemma~\ref{lemma:product}.

\begin{lemma}
\label{lemma:product}
Let $n \geq |\cX|$ and set $m = \lceil n/|\cX| \rceil$.  Suppose that $P[A^n|X^n]$ is a symmetric no-signalling conditional probability distribution.
Then $P[A^m|X^m]$ is separable, i.e., there exists a probability distribution $p_{\lambda_1, \ldots, \lambda_m}$ such that 
$$P[A^m|X^m]=\sum_{\lambda_1, \ldots, \lambda_m} p_{\lambda_1, \ldots, \lambda_m} P_{\lambda_1} [A_1|X_1] \cdots P_{\lambda_m} [A_m|X_m],$$
where $p_{\lambda_1, \ldots, \lambda_m}$ is a probability distribution on the labels $\lambda_1, \ldots, \lambda_m$, where $\lambda_j$ labels a finite set of conditional probability distributions.
\end{lemma}

\begin{proof}
  In order not to obscure the main argument, we prove the statement
  for integral $m = n/|\cX|$~\footnote{This immediately implies the result for $
  \lfloor n/|\cX| \rfloor$. The extension to the case $ \lceil n/|\cX|
  \rceil$ is more technical and can be found in~\cite{benmat:long}.}.
Our technique can be traced to
  Werner~\cite{werner89:_applic_inequal_quant_state_exten_probl}.
  We imagine the $m$ particles to be
  separated in space and note that $P[A^m|X^m]$ is separable if and only if it can be
  simulated by a local hidden variable model. Such a simulation is described in Fig.~1.
\begin{figure*}[tb]
\begin{center}
\includegraphics[]{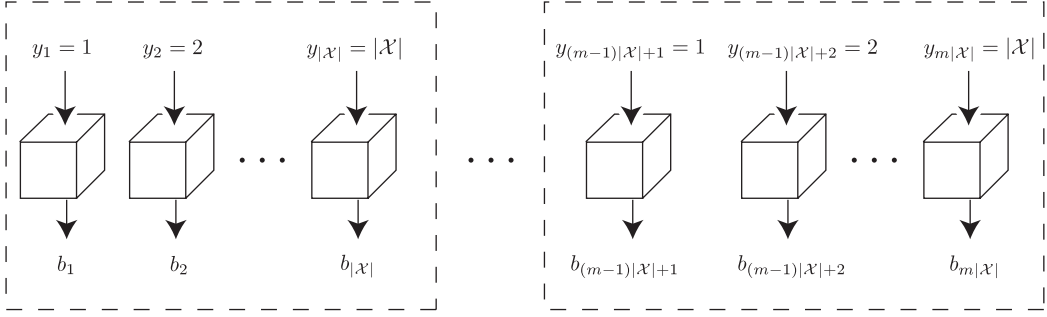}
\caption{Since $n = m |\cX|$, we can divide the particles into $m$
  groups of $|\cX|$ particles. In each of these groups we measure one
  particle according to each measurement in $X$ \emph{in advance} and
  record a list of all the results. In the simulation, if particle $i$ is supposed
  to be measured according to a measurement $x \in X$, we just look
  through the $i$th group until we come to the particle on which measurement $x$
  was performed in advance, and output the result we find.}
\end{center}\label{fig-boxes}
\end{figure*}
We now provide the formal proof. We construct a separable
conditional distribution
  $Q[A^m|X^m]$ and then show that it is equal to $P[A^m|X^m]$.  We assume that $\cX = \{1, 2, \ldots, |\cX|\}$, define a vector $y^n = (y_j)_{j=1,\ldots,n}$ with coordinates $y_j = (j-1 \mod |\cX| )+ 1$, and define the separable state
\begin{align*}
Q[A^m|X^m] = \sum_{b^n} q_{b^n} Q_{b^n, 1}[A_1|X_1] \cdots Q_{b^n,m}[A_m|X_m],
\end{align*}
where $b^n \in A^n$ is distributed according to $
  q_{b^n} = P[A^{n} = b^n |X^{n} = y^n]$
and the single-particle conditional probability distributions are
deterministic and defined by $Q_{b^n,i}[A_i = a_i|X_i = x_i] = [a_i = b_{(i-1)|\cX|+x_i}]$, where $[t]=1$ if $t$ is true and $0$ otherwise.
Let $\cL = \{1, 2, \ldots, n\}$, $\cL_1= \{(i-1)|\cX|+x_{i}: i = 1, 2, \ldots, m\}$ and $\cL_2 = \cL\backslash \cL_1$. Further let $A^{\cL}=A^n$, $A^{\cL_1}=(A_{x_1}, A_{|\cX|+x_2}, \ldots, A_{(m-1)|\cX|+x_m})$ and $A^{\cL_2}=A^{\cL} \backslash A^{\cL_1}$ and define $b^{\cL}, b^{\cL_1}$ and $b^{\cL_2}$ similarly. We find
\begin{align}
  Q&[A^m = a^m|X^m=x^m] \nonumber \\
  &=  \sum_{b^n} P[A^n=b^n|X^n=y^n] [a_1 = b_{x_1}]\cdots [a_{m} = b_{(m-1)|\cX|+x_{m}}] \nonumber\\
  & =   \sum_{b^{\cL_2}} P[A^{\cL_1}=a^m, A^{\cL_2}=b^{\cL_2} |X^{\cL_1}=x^m, X^{\cL_2}=y^{\cL_2}] \nonumber \\
    & =    P[A^{\cL_1}=a^m|X^{\cL_1}=x^m]  =  P[A^m=a^m|X^m=x^m],\nonumber
\end{align}
where we started with the definition of $Q[A^m|X^m]$, split the summation over $\cL_1$ and $\cL_2$, dropped the conditioning over $X^{\cL_2}=y^{\cL_2}$ because of the no-signalling property of $P$, used the definition of a marginal state, and, lastly, the permutation-invariance of $P$.
\end{proof}

Our next statement is a de Finetti theorem for symmetric separable conditional probability distributions.

\begin{lemma}\label{th:convex}
Suppose that $P[A^m|X^m]$ is a symmetric separable conditional probability distribution. Then there exists a probability distribution $p_\lambda$ such that 
\begin{align}
\begin{split}
\bigl\|P[A^k|X^k]- & \sum_\lambda p_\lambda P_\lambda[A|X]^{\times k} \bigr\|  \leq
 \min \biggl( \frac{2k|\cA|^{|\cX|}}{m}, \frac{k(k-1)}{m}\biggr),
\end{split}
\end{align}
where $p_\lambda$ is a probability distribution on a finite set of conditional probability distributions, labeled by $\lambda$.
\end{lemma}

\begin{proof}%
Let $Q_{1}[A|X], \ldots, Q_{E}[A|X]$ be the extreme points of the set of conditional probability distributions of one system. These are the deterministic functions $\cX \mapsto \cA$, hence $E=|\cA|^{|\cX|}$. Any symmetric separable conditional probability distribution is a convex combination of conditional probability distributions of the form $Q[A^m|X^m]=\frac{1}{m!} \sum_\pi Q_{i_{\pi^{-1}(1)}}[A|X]  \cdots Q_{i_{\pi^{-1}(m)}}[A|X] $, where $1\leq i_1, \ldots, i_m \leq E$.
%If $n$ is smaller than $E$ we may assume that only
%$n$ of the $\tau_i$ are different.
Define $Q[A|X]\coleq \frac{1}{m}
\sum_{j=1}^m Q_{i_j}[A|X]$. We expand
\begin{align}
 Q[A|X]^{\times k} =  \sum_{j_1=1}^m \cdots \sum_{j_k = 1}^m M_m(i_{j_1}, \ldots, i_{j_k}) Q_{i_{j_1}}[A_1|X_1]  \cdots  Q_{i_{j_k}}[A_m|X_m],
\end{align}
where $M_m(i_{j_1}, \ldots, i_{j_k}) = 1/m^k$ is the multinomial
distribution. To compare this expression with $Q[A^k|X^k]$,
write
\begin{align}
Q[A^k|X^k]=  \sum_{j_1=1}^m \cdots \sum_{j_k = 1}^m H_m(i_{j_1}, \ldots, i_{j_k}) Q_{i_{j_1}}[A_1|X_1] \cdots Q_{i_{j_k}}[A_m|X_m],
\end{align}
where $H_m(i_{j_1}, \ldots, i_{j_k})$ is the hypergeometric distribution for
an urn with $m$ balls (see~\cite{diaconis80:_finit}). Then
\begin{align}
\left\|  Q[A^k|X^k] -  Q[A|X]^{\times k}  \right\| &= \big\| \sum_{j_1,\ldots,j_k} \big( H_m(i_{j_1}, \ldots, i_{j_k})
 \nonumber\\ & \qquad M_m(i_{j_1}, \ldots, i_{j_k})  \big)Q_{i_{j_1}}[A_1|X_1] \cdots Q_{i_{j_k}}[A_m|X_m]\big\|  \nonumber\\ &\hspace{-1cm}\leq
\sum_{j_1,\ldots,j_k} \big| H_m(i_{j_1}, \ldots, i_{j_k})- M_m(i_{j_1}, \ldots, i_{j_k}) \big|\nonumber\\
& \hspace{-1cm}\leq \min \left( \frac{2kE}{m}, \frac{k(k-1)}{m}\right),
\end{align}
where we used the triangle inequality and Diaconis and
Freedman's result on estimating the hypergeometric distribution with
a multinomial distribution~\cite{diaconis80:_finit}.
\end{proof}

These two lemmas enable the proof of Theorem~\ref{theorem:1}.
\begin{proof}[Proof of Theorem~\ref{theorem:1}] Set $m = \lceil n/|\cX| \rceil$ and apply Lemma~\ref{lemma:product}. Then apply Lemma~\ref{th:convex}.
\end{proof}

Our final result is an application to quantum theory. In complete analogy to Lemma~\ref{th:convex} we show that the $k$-particle reduced state of a every separable symmetric density operator on $m$ copies of $\complex^d$ is approximated by a convex combination of tensor product states. Importantly, the approximation guarantee is independent of the dimension $d$, in contrast to the case of entangled states where a dependence on the dimension is necessary~\cite{christandl06:_one}. The norm is given by the trace norm $||A||_1=\tr \sqrt{A^\dagger A}$ for operators $A$ on $\complex^d$. It induces a distance measure on the set of quantum states that has a similar interpretation as a measure of distinguishability as the variational distance for probability distributions and the trace distance introduced on conditional probability distributions.
\begin{theorem}
If $\rho$ is a separable permutation-invariant density operator on
$(\complex^d)^{\otimes n}$, then there is a measure $m(\sigma)$ on states $\sigma$ on $\CC^d$ such that
\begin{align}
\bigl\|\rho^k- \int dm(\sigma) \, \sigma^{\otimes k} \bigr\|_1\leq
2\frac{k(k-1)}{n}.
\end{align}
\end{theorem}

\begin{proof}%
Any symmetric separable state is a convex combination of
states of the form $\omega^n=\frac{1}{n!} \sum_\pi \tau_{{\pi^{-1}(1)}} \otimes \cdots
\otimes \tau_{{\pi^{-1}(n)}}$, where $\{\tau_{j}\}_{j=1}^n$ is a set of pure states (these are extreme points in $\cB(\complex^d)$).
%If $n$ is smaller than $E$ we may assume that only
%$n$ of the $\tau_i$ are different.
Define $\tau\coleq \frac{1}{n}
\sum_{j=1}^n \tau_{j}$. We expand
\begin{align}
 \tau^{\otimes k} =  \sum_{j_1=1}^n \cdots \sum_{j_k = 1}^n M_n({j_1}, \ldots, {j_k}) \tau_{{j_1}} \otimes \cdots \otimes \tau_{{j_k}},
\end{align}
where $M_n({j_1}, \ldots, {j_k}) = 1/n^k$ is the multinomial
distribution. To compare this expression with $\omega^k:=\tr_{n-k} \omega^n$,
write
\begin{align}
 \omega^k =  \sum_{j_1=1}^n \cdots \sum_{j_k = 1}^n H_n({j_1}, \ldots, {j_k}) \tau_{{j_1}} \otimes \cdots \otimes \tau_{{j_k}},
\end{align}
where $H_n({j_1}, \ldots, {j_k})$ is the hypergeometric distribution for
an urn with $n$ balls (see~\cite{diaconis80:_finit}). Then
\begin{align}
\left\| \omega^k - \tau^{\otimes k} \right\|_1 &= \big\| \sum_{j_1,\ldots,j_k} \big( H_n({j_1}, \ldots, {j_k})
 \nonumber\\ & \qquad M_n({j_1}, \ldots, {j_k})  \big)\tau_{{j_1}} \otimes \cdots \otimes \tau_{{j_k}}\big\|_1  \nonumber\\ &\hspace{-1cm}\leq
\sum_{j_1,\ldots,j_k} \big| H_n({j_1}, \ldots, {j_k})- M_n({j_1}, \ldots, {j_k}) \big|\nonumber\\
& \hspace{-1cm}\leq \frac{k(k-1)}{n},
\end{align}
where we used the triangle inequality and Diaconis and
Freedman's result on estimating the hypergeometric distribution with
a multinomial distribution~\cite{diaconis80:_finit}.
\end{proof}

\section{Towards a finite de Finetti theorem for the convex sets framework}

\label{sec:conv-sets-fram}
We will start this section with a self-contained introduction to the convex sets framework. (See Refs.~\cite{Barnum:06} and~\cite{Barnum:07} for a gentler introduction.) We will then generalise Lemma~\ref{th:convex} to this setting. Finally, we pose the question of the existence of a finite de Finetti theorem in the convex sets framework.

Let $\Omega$ be the set of states of a particle.
We assume that $\Omega$ is convex, compact, and has affine dimension $n$. In probability theory, for example, $\Omega$ is the simplex of
probability distributions $(\omega_1, \ldots, \omega_{n+1})$, $\omega_i \geq 0, \sum_i \omega_i=1$, while in quantum theory, $\Omega$ is (isomorphic to) the set of positive
operators $\omega$ with trace one on a Hilbert space $\cH\cong \complex^d$. We are particularly interested in the case where $\Omega$ is specified by a set of conditional probability distributions $\{ P_\lambda[A|X]\}$, whose elements are indexed by a label $\lambda$.
 This is partly because quantum states can be described in this way. For instance, the state $\rho$ of a qubit, a spin-$\frac{1}{2}$ system, is uniquely determined by the probabilities of obtaining spin up or down when it is measured along the $x$, $y$, or $z$ axes of the Bloch sphere. Thus a qubit can be described by a conditional probability distribution $P[A|X]$ with $\cA = \{ \uparrow, \downarrow\}$ and $\cX = \{x,y,z\}$. % \bnote{I would stop paragraph here.} $P[\uparrow|\mbox{axis}=x]=\tr \proj{\uparrow_x} \rho, P[\downarrow|\mbox{axis}=x]=\tr \proj{\downarrow_x} \rho$, $P[\uparrow|\mbox{axis}=y]=\tr \proj{\uparrow_y} \rho$ and so on, where $\ket{\uparrow_{i}}$ and $\ket{\downarrow_{i}}$ are the eigenvectors of the Pauli matrix $\sigma_i$ with eigenvalue $\pm 1$.
Not all conditional probability distributions can be obtained by making local measurements on quantum states. This led Barrett to define generalized theories~\cite{Barrett:05}, where the state space $\Omega$ is the set of all conditional probability distributions $\{P_\lambda[A|X]\}$, denoted $\boxy$. This is the case that we considered in the previous parts of the paper. When $|\cX| = 1$, this reduces to classical probability theory.
  In quantum theory, $|\cX| = 1$ corresponds to the case where all measurements on a system commute, and thus can be performed at once.
 In fact, every $\Omega$ can be mapped to a
convex subset of $\boxy$ for some number of \emph{fiducial} measurements and
outcomes~\cite[Lemma~1]{Barnum:06}.

In quantum theory, the most general measurement
that can be performed is a positive operator-valued measure (POVM),
whose elements are termed \emph{effects}. Effects are linear functions
mapping states to probabilities: in (finite-dimensional) quantum theory, the probability of
obtaining the outcome associated with an effect $r$, when the state is
$\omega$, is $r(\omega)=\tr \left(R \omega\right)$ for some bounded nonnegative operator $R$
with $R \leq \mathbf{1}$.
In a generalized theory, effects are also functions mapping states to probabilities, and these functions should be affine so that they are compatible with preparing convex combinations.
The vector space of affine functions $a:
\Omega \to \RR$, denoted $A(\Omega)$, is isomorphic to $\RR^{n+1}$.
% The elements of $A(\Omega)$ can be
% partially ordered via the relation \bnote{delete this} $a\geq b$ iff $a(\omega)\geq
% b(\omega)$ for all
% $\omega \in \Omega$.
The cone of nonnegative affine functions on $\Omega$ is denoted $A_+(\Omega)$.
The \emph{order unit} of $A(\Omega)$ is the
element $e \in A(\Omega)$ satisfying $e(\omega)=1$ for all $\omega \in \Omega$. An
\emph{effect} is an element $a\in A(\Omega)$ satisfying $0\leq a(\omega)
\leq 1$ for all $\omega \in \Omega$. % The real number $a(\omega)$ is accordingly the probability of the event
% associated with the effect $a$.
The set of all effects is denoted $[0, e]$.
There is a natural embedding of $\Omega$ into $A(\Omega)^*$, the dual space of $A(\Omega)$, given by $\omega \mapsto \hat \omega$, where $\hat \omega (a) = a(\omega) $ for all $a \in A(\Omega)$. Furthermore, if $\hat \omega \in A(\Omega)^*$ satisfies $\hat \omega(a) \geq 0$ for all $a \in A_+(\Omega)$ and $\hat \omega(e) = 1$, then $\hat \omega$ is the image of some state $\omega \in \Omega$~\cite[Section~2.6]{Boyd:04}. We identify $\hat \omega$ with $\omega$ in what follows. It is easy to check that $\|\cdot \|\:=\sup_{a \in [0, e]} |a(\cdot)|$ is a norm on $A(\Omega)^*$. For more details about the convex sets framework, see~\cite{Barnum:06,Barnum:07}.

A natural distance measure on the set of states, which  generalises
the variational distance between classical probability distributions and the trace distance between quantum states, is given by
\begin{align} \label{eq:distance}\|\omega - \omega' \|\:=\sup_{a \in [0, e]} |a(\omega)-a(\omega')|.\end{align}

In quantum theory, systems are combined by taking the \emph{tensor product} of the Hilbert spaces for each system. The same is true in the convex sets framework: $\omega \otimes \omega'$ is defined to be the \emph{product state} where system $\Omega$ is in state $\omega$, system $\Omega'$ is in state $\omega'$, and the two systems are independent. The complication is that the space $A(\Omega)^\star$ is a Banach space but not a Hilbert space and there are multiple ways to define a norm on the tensor product space, consistent with the norm on $A(\Omega)^\star$. This choice affects the set of pure (i.e., norm $1$) states of the joint system.
At the very least, we want the set of joint states to be closed under convex combinations. This yields:

\begin{definition}
The \emph{minimal tensor product} of $\Omega$ and $\Omega'$, denoted
by $\Omega\tmin \Omega'$ consists of all convex combinations of product states $\omega \otimes \omega'$, $\omega \in \Omega$ and $\omega' \in \Omega'$.
\end{definition}

We say that states in $\Omega \tmin \Omega'$ are \emph{separable}, thereby extending terminology from quantum mechanics to the convex sets framework.
Next, if $a$ is a valid effect for system $\Omega$ and $a'$ a valid effect for system $\Omega'$, then $a \otimes a'$ is the effect defined on product states via $a \otimes a'(\omega \otimes \omega') = a(\omega)a'(\omega')$. If all convex combinations of such effects are to be allowed, the state space must only contain states in the \emph{maximal tensor product}, defined via duality as:

\begin{definition}
The \emph{maximal tensor product} of $\Omega$ and $\Omega'$, denoted
by $\Omega\tmax \Omega'$ consists of all bilinear functions $\mu:
A(\Omega)\times A(\Omega') \rightarrow \mathbb{R}$ that satisfy $\mu
(a \otimes b)\geq 0$ for $a, b \geq 0$, and $\mu(e \otimes e')=1$.
\end{definition}

Thus $\mu \in \Omega \tmax \Omega'$ can be written as a linear
combination of product states, possibly with negative weights. In
classical probability theory, the minimal and the maximal tensor
product coincide. In general, a tensor product $\Omega \otimes
\Omega'$ is a convex set with $\Omega \tmin \Omega'\subseteq \Omega
\otimes \Omega' \subseteq \Omega \tmax \Omega'$. In quantum theory,
$\Omega \otimes \Omega'$ is the set of trace one positive operators on
the (unique) Hilbert space tensor product of $\cH$ and $\cH'$. Note
that $\Omega \otimes \Omega'$ lies strictly between the maximal and
minimal tensor products in the quantum case. The set of separable quantum states is
$\Omega \otimes_\text{min} \Omega'$ and $\Omega \otimes_\text{max} \Omega'$ is the set of trace one entanglement witnesses.

For a state $\mu\in \Omega
\otimes \Omega'$, we say that $\mu_{\Omega} \in \Omega$, defined by
$a(\mu_{\Omega})=a \otimes e' (\mu)$ for all effects $a$, is the
\emph{partial trace} of $\mu$ with respect to $\Omega'$. An effect on
the tensor product is an element $a \in A(\Omega \otimes \Omega')$
satisfying $0 \leq a \leq e \otimes e'$. The larger the set of joint
states, the smaller the set of allowed effects.  This means that the
distance measure that we defined in Eq.~(\ref{eq:distance}), when
applied to states of more than one particle, depends on which tensor
product we use. It is true, however, that $\|\omega - \omega'\| \leq
\|\omega -\omega'\|_\text{min}$, the distance measure for the minimal
tensor product, since in that case the set of effects is largest.
Also note that a physical theory may place additional restrictions on
which effects are allowed but, even then, $\|\omega - \omega'\|$
provides an upper bound on the probability of distinguishing $\omega$
and $\omega'$.

\begin{theorem}\label{th:convex-gen}
Let $\Omega$ be a convex set with $E$ extreme points ($E$ may be infinite). Suppose
$\omega^n \in \Omega^{\tmin n}$ is symmetric. Then there is a
measure $m(\tau)$ on states $\tau \in \Omega$ such that
\begin{align}
\bigl\|\omega^k- \int dm(\tau) \, \tau^{\otimes k}\bigr\|_\text{min}\leq \min \left(
\frac{2kE}{n}, \frac{k(k-1)}{n}\right).
\end{align}
\end{theorem}

\begin{proof}%
Let $\tau_1, \ldots, \tau_E$ be the extreme points of $\Omega$.
Any symmetric separable state is a convex combination of
states of the form $\omega^n=\frac{1}{n!} \sum_\pi \tau_{i_{\pi^{-1}(1)}} \otimes \cdots
\otimes \tau_{i_{\pi^{-1}(n)}}$, where $1\leq i_1, \ldots, i_n \leq E$.
%If $n$ is smaller than $E$ we may assume that only
%$n$ of the $\tau_i$ are different.
Define $\tau\coleq \frac{1}{n}
\sum_{j=1}^n \tau_{i_j}$. We expand
\begin{align}
 \tau^{\otimes k} =  \sum_{j_1=1}^n \cdots \sum_{j_k = 1}^n M_n(i_{j_1}, \ldots, i_{j_k}) \tau_{i_{j_1}} \otimes \cdots \otimes \tau_{i_{j_k}},
\end{align}
where $M_n(i_{j_1}, \ldots, i_{j_k}) = 1/n^k$ is the multinomial
distribution. To compare this expression with $\omega^k$,
write
\begin{align}
 \omega^k =  \sum_{j_1=1}^n \cdots \sum_{j_k = 1}^n H_n(i_{j_1}, \ldots, i_{j_k}) \tau_{i_{j_1}} \otimes \cdots \otimes \tau_{i_{j_k}},
\end{align}
where $H_n(i_{j_1}, \ldots, i_{j_k})$ is the hypergeometric distribution for
an urn with $n$ balls (see~\cite{diaconis80:_finit}). Then
\begin{align}
\left\| \omega^k - \tau^{\otimes k} \right\|_\text{min} &= \big\| \sum_{j_1,\ldots,j_k} \big( H_n(i_{j_1}, \ldots, i_{j_k})
 \nonumber\\ & \qquad M_n(i_{j_1}, \ldots, i_{j_k})  \big)\tau_{i_{j_1}} \otimes \cdots \otimes \tau_{i_{j_k}}\big\|_\text{min}  \nonumber\\ &\hspace{-1cm}\leq
\sum_{j_1,\ldots,j_k} \big| H_n(i_{j_1}, \ldots, i_{j_k})- M_n(i_{j_1}, \ldots, i_{j_k}) \big|\nonumber\\
& \hspace{-1cm}\leq \min \left( \frac{2kE}{n}, \frac{k(k-1)}{n}\right),
\end{align}
where we used the triangle inequality and Diaconis and
Freedman's result on estimating the hypergeometric distribution with
a multinomial distribution~\cite{diaconis80:_finit}.
\end{proof}

One can show that $\boxy^{\tmax n}$ is precisely the set of all no-signalling conditional probability distributions and that $\boxy^{\tmin n}$ is the set of all separable conditional probability distributions ~\cite{randall81:_operat_statis_tensor_produc,Barrett:05}. Furthermore the trace distance (Definition~\ref{def:tracedistance}) coincides with the definition in Eq.~(\ref{eq:distance}). With these observations and the fact that $||\cdot ||\leq || \cdot ||_{\min}$ we see that Theorem~\ref{th:convex-gen} generalises Lemma~\ref{th:convex}. Unfortunately, we were not able to obtain a similar generalisation of Lemma~\ref{lemma:product} and hence of Theorem~\ref{theorem:1}. We thus conclude with the question of whether a finite de Finetti theorem exists for general theories in the convex sets framework. We remark that the argument of~\cite{caves:4537} applied in this context yields an infinite de Finetti theorem for any theory in the convex sets framework (see~\cite{BarrettLeifer06} for the details).

% \emph{Acknowledgements.}---This work was carried out at the same time as related work by J.~Barrett and M.~Leifer~\cite{BarrettLeifer06}. We thank them for
% discussions of their work, and especially for explaining how to define the analogue of trace distance. We also thank R.~Colbeck and R.~Renner for discussions, G.~Mitchison for valuable comments on the manuscript, and the organizers of the FQXi workshop \emph{Operational probabilistic theories as foils to quantum theory}, where part of this work was done. MC would like to thank the IQI at Caltech
% and the CWI in Amsterdam for their hospitality and acknowledge the Research Fellowships of the UK's EPSRC and Magdalene College Cambridge. This work was supported in part by the
% NSF under Grants PHY-0456720 and CCF-0524828,
% the EU's FP6-FET Integrated Projects SCALA (CT-015714) and QAP
% (CT-015848), NWO VICI project 639-023-302, and the Dutch BSIK/BRICKS project.

\section{Acknowledgements}
This work was carried out at the same time as related work by J.~Barrett and M.~Leifer~\cite{BarrettLeifer06}. We thank them for
discussions, and especially for explaining how to define the trace
distance. We thank R.~Colbeck and R.~Renner for discussions,
G.~Mitchison for valuable comments on the manuscript, and the organizers of
the FQXi workshop \emph{Operational probabilistic theories as foils to
  quantum theory}, where part of this work was done.  MC thanks the
IQI at Caltech
and CWI Amsterdam for their hospitality.
 This work was supported by the UK's EPSRC, Magdalene College Cambridge, NSF Grants PHY-0456720 and CCF-0524828,
EU Projects SCALA (CT-015714) and QAP
(CT-015848), NWO VICI project 639-023-302, and the Dutch BSIK/BRICKS project.

%  \bibliography{newref}

\begin{thebibliography}{27}
\expandafter\ifx\csname natexlab\endcsname\relax\def\natexlab#1{#1}\fi
\expandafter\ifx\csname bibnamefont\endcsname\relax
  \def\bibnamefont#1{#1}\fi
\expandafter\ifx\csname bibfnamefont\endcsname\relax
  \def\bibfnamefont#1{#1}\fi
\expandafter\ifx\csname citenamefont\endcsname\relax
  \def\citenamefont#1{#1}\fi
\expandafter\ifx\csname url\endcsname\relax
  \def\url#1{\texttt{#1}}\fi
\expandafter\ifx\csname urlprefix\endcsname\relax\def\urlprefix{URL }\fi
\providecommand{\bibinfo}[2]{#2}
\providecommand{\eprint}[2][]{\url{#2}}



\bibitem[{\citenamefont{Diaconis and Freedman}(1980)}]{diaconis80:_finit}
\bibinfo{author}{\bibfnamefont{P.}~\bibnamefont{Diaconis}} \bibnamefont{and}
  \bibinfo{author}{\bibfnamefont{D.}~\bibnamefont{Freedman}},
  \bibinfo{journal}{Ann. Probab.} \textbf{\bibinfo{volume}{8}},
  \bibinfo{pages}{745} (\bibinfo{year}{1980}).

\bibitem[{\citenamefont{Bernardo and Smith}(1994)}]{bernardo94:_bayes_theor}
\bibinfo{author}{\bibfnamefont{J.~M.} \bibnamefont{Bernardo}} \bibnamefont{and}
  \bibinfo{author}{\bibfnamefont{A.~F.~M.} \bibnamefont{Smith}},
  \emph{\bibinfo{title}{Bayesian Theory}} (\bibinfo{publisher}{Wiley},
  \bibinfo{address}{Chichester}, \bibinfo{year}{1994}).

\bibitem[{\citenamefont{{de Finetti}}(1937)}]{finetti37:_la}
\bibinfo{author}{\bibfnamefont{B.}~\bibnamefont{{de Finetti}}},
  \bibinfo{journal}{Ann. Inst. H. Poincar{\'{e}}} \textbf{\bibinfo{volume}{7}},
  \bibinfo{pages}{1} (\bibinfo{year}{1937}).

\bibitem[{\citenamefont{K{\"{o}}nig and Renner}(2005)}]{Koenig:04a}
\bibinfo{author}{\bibfnamefont{R.}~\bibnamefont{K{\"{o}}nig}} \bibnamefont{and}
  \bibinfo{author}{\bibfnamefont{R.}~\bibnamefont{Renner}},
  \bibinfo{journal}{J. Math. Phys.} \textbf{\bibinfo{volume}{46}},
  \bibinfo{eid}{122108} (\bibinfo{year}{2005}),
  \bibinfo{note}{quant-ph/0410229}.

\bibitem[{\citenamefont{Christandl et~al.}(2007)\citenamefont{Christandl,
  K{\"o}nig, Mitchison, and Renner}}]{christandl06:_one}
\bibinfo{author}{\bibfnamefont{M.}~\bibnamefont{Christandl}},
  \bibinfo{author}{\bibfnamefont{R.}~\bibnamefont{K{\"o}nig}},
  \bibinfo{author}{\bibfnamefont{G.}~\bibnamefont{Mitchison}},
  \bibnamefont{and} \bibinfo{author}{\bibfnamefont{R.}~\bibnamefont{Renner}},
  \bibinfo{journal}{Comm. Math. Phys.} \textbf{\bibinfo{volume}{273}},
  \bibinfo{pages}{473} (\bibinfo{year}{2007}),
  \bibinfo{note}{quant-ph/0602130}.

\bibitem[{\citenamefont{Renner}(2005)}]{renner05:_secur}
\bibinfo{author}{\bibfnamefont{R.}~\bibnamefont{Renner}}, Ph.D. thesis,
  \bibinfo{school}{Swiss Federal Institute of Technology},
  \bibinfo{address}{Z{\"u}rich} (\bibinfo{year}{2005}),
  \bibinfo{note}{quant-ph/0512258};
%\bibitem[{\citenamefont{Renner}(2007)}]{renner:nature}
%\bibinfo{author}{\bibfnamefont{R.}~\bibnamefont{Renner}},
  \bibinfo{journal}{Nature Physics} \textbf{\bibinfo{volume}{3}},
  \bibinfo{pages}{645 } (\bibinfo{year}{2007}),
  \bibinfo{note}{quant-ph/0703069}.

\bibitem[{\citenamefont{Bennett and Brassard}(1984)}]{QKD}
\bibinfo{author}{\bibfnamefont{C.~H.} \bibnamefont{Bennett}} \bibnamefont{and}
  \bibinfo{author}{\bibfnamefont{G.}~\bibnamefont{Brassard}}, in
  \emph{\bibinfo{booktitle}{Proceedings of IEEE International Conference on
  Computers, Systems, and Signal Processing}} (\bibinfo{publisher}{IEEE},
  \bibinfo{year}{1984}), pp. \bibinfo{pages}{175--179};
%\bibitem[{\citenamefont{Ekert}(1991)}]{PhysRevLett.67.661}
\bibinfo{author}{\bibfnamefont{A.~K.} \bibnamefont{Ekert}},
  \bibinfo{journal}{Phys. Rev. Lett.} \textbf{\bibinfo{volume}{67}},
  \bibinfo{pages}{661} (\bibinfo{year}{1991}).


\bibitem[{\citenamefont{Ac\'{\i}n et~al.}(2007)\citenamefont{Ac\'{\i}n,
 Brunner, Gisin, Massar, Pironio, and Scarani}}]{acin:230501}
\bibinfo{author}{\bibfnamefont{A.}~\bibnamefont{Ac\'{\i}n}},
 \bibinfo{author}{\bibfnamefont{N.}~\bibnamefont{Brunner}},
 \bibinfo{author}{\bibfnamefont{N.}~\bibnamefont{Gisin}},
 \bibinfo{author}{\bibfnamefont{S.}~\bibnamefont{Massar}},
 \bibinfo{author}{\bibfnamefont{S.}~\bibnamefont{Pironio}}, \bibnamefont{and}
 \bibinfo{author}{\bibfnamefont{V.}~\bibnamefont{Scarani}},
 \bibinfo{journal}{Phys. Rev. Lett.} \textbf{\bibinfo{volume}{98}},
 \bibinfo{eid}{230501} (\bibinfo{year}{2007}),
 \bibinfo{note}{quant-ph/0702152};
%\bibitem[{\citenamefont{Barrett et~al.}(2005)\citenamefont{Barrett, Hardy, and  Kent}}]{Barrett:05a}
\bibinfo{author}{\bibfnamefont{J.}~\bibnamefont{Barrett}},
  \bibinfo{author}{\bibfnamefont{L.}~\bibnamefont{Hardy}}, \bibnamefont{and}
  \bibinfo{author}{\bibfnamefont{A.}~\bibnamefont{Kent}},
  \bibinfo{journal}{Phys. Rev. Lett.} \textbf{\bibinfo{volume}{95}},
  \bibinfo{pages}{010503} (\bibinfo{year}{2005}),
  \bibinfo{note}{quant-ph/0405101};
%\bibitem[{\citenamefont{{L}l. Masanes and Winter}()}]{masanes:_uncon}
\bibinfo{author}{\bibnamefont{{L}l. Masanes}},   \bibinfo{author}{\bibfnamefont{R.}~\bibnamefont{Renner}},
  \bibinfo{author}{\bibfnamefont{A.}~\bibnamefont{Winter}},   \bibinfo{author}{\bibfnamefont{J.}~\bibnamefont{Barrett}}   \bibnamefont{and}   \bibinfo{author}{\bibfnamefont{M.}~\bibnamefont{Christandl}},
  \bibinfo{note}{quant-ph/0606049};
%\bibitem[{\citenamefont{Ac{\'\i}n et~al.}(2006)\citenamefont{Ac{\'\i}n, Gisin, and {L}l. Masanes}}]{acin05:_from_theor_secur_quant_key_distr}
\bibinfo{author}{\bibfnamefont{A.}~\bibnamefont{Ac{\'\i}n}},
  \bibinfo{author}{\bibfnamefont{N.}~\bibnamefont{Gisin}}, \bibnamefont{and}
  \bibinfo{author}{\bibnamefont{{L}l. Masanes}}, \bibinfo{journal}{Phys. Rev.
  Lett.} \textbf{\bibinfo{volume}{97}}, \bibinfo{eid}{120405}
  (\bibinfo{year}{2006}), \bibinfo{note}{quant-ph/0510094};
%\bibitem[{\citenamefont{Scarani et~al.}(2006)\citenamefont{Scarani, Gisin,  Brunner, {L}l. Masanes, Pino, and Ac{\'\i}n}}]{scarani:_secrec}
\bibinfo{author}{\bibfnamefont{V.}~\bibnamefont{Scarani}},
  \bibinfo{author}{\bibfnamefont{N.}~\bibnamefont{Gisin}},
  \bibinfo{author}{\bibfnamefont{N.}~\bibnamefont{Brunner}},
  \bibinfo{author}{\bibnamefont{{L}l. Masanes}},
  \bibinfo{author}{\bibfnamefont{S.}~\bibnamefont{Pino}}, \bibnamefont{and}
  \bibinfo{author}{\bibfnamefont{A.}~\bibnamefont{Ac{\'\i}n}},
  \bibinfo{journal}{Phys. Rev. A} \textbf{\bibinfo{volume}{74}},
  \bibinfo{eid}{042339} (\bibinfo{year}{2006}),
  \bibinfo{note}{quant-ph/0606197}.

\bibitem[{\citenamefont{Caves et~al.}(2002)\citenamefont{Caves, Fuchs, and
  Schack}}]{caves:4537}
\bibinfo{author}{\bibfnamefont{C.~M.} \bibnamefont{Caves}},
  \bibinfo{author}{\bibfnamefont{C.~A.} \bibnamefont{Fuchs}}, \bibnamefont{and}
  \bibinfo{author}{\bibfnamefont{R.}~\bibnamefont{Schack}},
  \bibinfo{journal}{J. Math. Phys.} \textbf{\bibinfo{volume}{43}},
  \bibinfo{pages}{4537} (\bibinfo{year}{2002}),
  \bibinfo{note}{quant-ph/0104088}.

\bibitem[{\citenamefont{Barnum et~al.}(2006)\citenamefont{Barnum, Barrett,
  Leifer, and Wilce}}]{Barnum:06}
\bibinfo{author}{\bibfnamefont{H.}~\bibnamefont{Barnum}},
  \bibinfo{author}{\bibfnamefont{J.}~\bibnamefont{Barrett}},
  \bibinfo{author}{\bibfnamefont{M.}~\bibnamefont{Leifer}}, \bibnamefont{and}
  \bibinfo{author}{\bibfnamefont{A.}~\bibnamefont{Wilce}}
 (\bibinfo{year}{2007}), \bibinfo{note}{quant-ph/0611295}.

\bibitem[{\citenamefont{Barnum et~al.}(2007)\citenamefont{Barnum, Barrett,
  Leifer, and Wilce}}]{Barnum:07}
\bibinfo{author}{\bibfnamefont{H.}~\bibnamefont{Barnum}},
  \bibinfo{author}{\bibfnamefont{J.}~\bibnamefont{Barrett}},
  \bibinfo{author}{\bibfnamefont{M.}~\bibnamefont{Leifer}}, \bibnamefont{and}
  \bibinfo{author}{\bibfnamefont{A.}~\bibnamefont{Wilce}}, \bibinfo{journal}{Phys. Rev. Lett.} \textbf{\bibinfo{volume}{99}}, \bibinfo{pages}{240501}
  (\bibinfo{year}{2007}), \bibinfo{note}{arXiv:0707.0620}.

\bibitem[{\citenamefont{Barrett and Leifer}(2007)}]{BarrettLeifer06}
\bibinfo{author}{\bibfnamefont{J.}~\bibnamefont{Barrett}} \bibnamefont{and}
  \bibinfo{author}{\bibfnamefont{M.}~\bibnamefont{Leifer}}
  (\bibinfo{year}{2007}), \bibinfo{note}{arXiv:0712.2265}.


\bibitem[{\citenamefont{St{\o}rmer}(1969)}]{stormer69:_symmet_states_infin_ten%
sor_produc_c}
\bibinfo{author}{\bibfnamefont{E.}~\bibnamefont{St{\o}rmer}},
  \bibinfo{journal}{J. Funct. Anal.} \textbf{\bibinfo{volume}{3}},
  \bibinfo{pages}{48} (\bibinfo{year}{1969});
%\bibitem[{\citenamefont{Hudson and  Moody}(1976)}]{hudson76:_local_normal_symmet_states_analog_theor}
\bibinfo{author}{\bibfnamefont{R.~L.} \bibnamefont{Hudson}} \bibnamefont{and}
  \bibinfo{author}{\bibfnamefont{G.~R.} \bibnamefont{Moody}},
  \bibinfo{journal}{Z. Wahrschein. verw. Geb. (Probab. Theory Related Fields)}
  \textbf{\bibinfo{volume}{33}}, \bibinfo{pages}{343} (\bibinfo{year}{1976}).

\bibitem[{\citenamefont{Fuchs et~al.}(2004)\citenamefont{Fuchs, Schack, and
  Scudo}}]{fuchs:062305}
\bibinfo{author}{\bibfnamefont{C.~A.} \bibnamefont{Fuchs}},
  \bibinfo{author}{\bibfnamefont{R.}~\bibnamefont{Schack}}, \bibnamefont{and}
  \bibinfo{author}{\bibfnamefont{P.~F.} \bibnamefont{Scudo}},
  \bibinfo{journal}{Phys. Rev. A} \textbf{\bibinfo{volume}{69}},
  \bibinfo{eid}{062305} (\bibinfo{year}{2004}),
  \bibinfo{note}{quant-ph/0307198}.

\bibitem[{\citenamefont{Christandl and Toner}(2009)}]{benmat:long}
\bibinfo{author}{\bibfnamefont{M.}~\bibnamefont{Christandl}} \bibnamefont{and}
  \bibinfo{author}{\bibfnamefont{B.}~\bibnamefont{Toner}}
  (\bibinfo{year}{2009}), \bibinfo{note}{in preparation}.

\bibitem[{\citenamefont{Eggeling and Werner}(2002)}]{EggWer2002}
\bibinfo{author}{\bibfnamefont{T.}~\bibnamefont{Eggeling}} \bibnamefont{and}
  \bibinfo{author}{\bibfnamefont{R.}~\bibnamefont{Werner}},
  \bibinfo{journal}{Phys. Rev. Lett.} \textbf{\bibinfo{volume}{89}},
  \bibinfo{pages}{097905} (\bibinfo{year}{2002}).

\bibitem[{\citenamefont{Hayden et~al.}(2005)\citenamefont{Hayden, Leung, and
  Smith}}]{hayden:062339}
\bibinfo{author}{\bibfnamefont{P.}~\bibnamefont{Hayden}},
  \bibinfo{author}{\bibfnamefont{D.}~\bibnamefont{Leung}}, \bibnamefont{and}
  \bibinfo{author}{\bibfnamefont{G.}~\bibnamefont{Smith}},
  \bibinfo{journal}{Phys. Rev. A} \textbf{\bibinfo{volume}{71}},
  \bibinfo{eid}{062339} (\bibinfo{year}{2005}).

\bibitem[{\citenamefont{Barrett}(2007)}]{Barrett:05}
\bibinfo{author}{\bibfnamefont{J.}~\bibnamefont{Barrett}},
  \bibinfo{journal}{Phys. Rev. A} \textbf{\bibinfo{volume}{75}},
  \bibinfo{eid}{032304} (\bibinfo{year}{2007}),
  \bibinfo{note}{quant-ph/0508211}.

\bibitem[{\citenamefont{Werner}(1989)}]{werner89:_applic_inequal_quant_state_e%
xten_probl}
\bibinfo{author}{\bibfnamefont{R.~F.} \bibnamefont{Werner}},
  \bibinfo{journal}{Lett. Math. Phys.} \textbf{\bibinfo{volume}{17}},
  \bibinfo{pages}{359} (\bibinfo{year}{1989});
  %\bibitem[{\citenamefont{Terhal et~al.}(2003)\citenamefont{Terhal, Doherty, and Schwab}}]{Terhal:03a}
\bibinfo{author}{\bibfnamefont{B.~M.} \bibnamefont{Terhal}},
  \bibinfo{author}{\bibfnamefont{A.~C.} \bibnamefont{Doherty}},
  \bibnamefont{and} \bibinfo{author}{\bibfnamefont{D.}~\bibnamefont{Schwab}},
  \bibinfo{journal}{Phys. Rev. Lett.} \textbf{\bibinfo{volume}{90}},
  \bibinfo{pages}{157903} (\bibinfo{year}{2003}),
  \bibinfo{note}{quant-ph/0210053};
%\bibitem[{\citenamefont{Toner}(2006)}]{toner:_monog}
\bibinfo{author}{\bibfnamefont{B.~F.} \bibnamefont{Toner}},
  \bibinfo{journal}{Proc. R. Soc. A} \textbf{\bibinfo{volume}{465}},
  \bibinfo{pages}{59--69}  (\bibinfo{year}{2009}), \bibinfo{note}{quant-ph/0601172}.

\bibitem[{\citenamefont{Boyd and Vandenberghe}(2004)}]{Boyd:04}
\bibinfo{author}{\bibfnamefont{S.}~\bibnamefont{Boyd}} \bibnamefont{and}
  \bibinfo{author}{\bibfnamefont{L.}~\bibnamefont{Vandenberghe}},
  \emph{\bibinfo{title}{Convex Optimization}} (\bibinfo{publisher}{Cambridge
  University Press}, \bibinfo{address}{Cambridge}, \bibinfo{year}{2004}),
  \bibinfo{note}{available online at
  http://www.stanford.edu/\~{}boyd/cvxbook/}.

\bibitem[{\citenamefont{Randall and
  Foulis}(1981)}]{randall81:_operat_statis_tensor_produc}
\bibinfo{author}{\bibfnamefont{C.~H.} \bibnamefont{Randall}} \bibnamefont{and}
  \bibinfo{author}{\bibfnamefont{D.~J.} \bibnamefont{Foulis}}, in
  \emph{\bibinfo{booktitle}{Interpretations and Foundations of Quantum
  Mechanics}}, edited by
  \bibinfo{editor}{\bibfnamefont{H.}~\bibnamefont{Neumann}}
  (\bibinfo{publisher}{Bibliographisches Institut, Wissenschaftsverlag},
  \bibinfo{address}{Manheim}, \bibinfo{year}{1981}).

\bibitem[{\citenamefont{Popescu and Rohrlich}(1994)}]{Popescu:94a}
\bibinfo{author}{\bibfnamefont{S.}~\bibnamefont{Popescu}} \bibnamefont{and}
  \bibinfo{author}{\bibfnamefont{D.}~\bibnamefont{Rohrlich}},
  \bibinfo{journal}{Found. Phys.} \textbf{\bibinfo{volume}{24}},
  \bibinfo{pages}{379} (\bibinfo{year}{1994}).


\end{thebibliography}
%  \end{document}

\end{document}